\documentclass{elsarticle}               % The use of LaTeX2e is preferred.

\usepackage{amsmath,amssymb,amsfonts}
\usepackage{amsthm}
\usepackage{algorithmic}
\usepackage{graphicx}
\usepackage{float}
\usepackage{textcomp}

  \pdfoutput=1  % For arxiv

\newtheorem{propo}{Proposition}

\newtheorem{property}{Property}
 
\newtheorem{corollary}{Corollary}

\newtheorem{rema}{Remark}

\begin{document}

\begin{frontmatter}
%\runtitle{Insert a suggested running title}  % Running title for regular 
                                              % papers but only if the title  
                                              % is over 5 words. Running title 
                                              % is not shown in output.

\title{Stochastic  Event  Generation Through  Markovian Jumps: Generalized Distribution and Minimal Representations} % Title, preferably not more 
                                                % than 10 words.

\author{Carlos Andrey Maia\fnref{myfootnote}}
\address{Departamento de Engenharia El\'{e}trica, Universidade Federal de Minas Gerais, Av.
Ant\^{o}nio Carlos 6627, Pampulha, 31270-010, Belo Horizonte, MG,
BRAZIL.}
\fntext[myfootnote]{Corresponding author: maia@cpdee.ufmg.br }

%\author[a]{Carlos Andrey Maia}\ead{maia@cpdee.ufmg.br }    % Add the 
%
%
%\address[a]{Departamento de
%Engenharia El\'{e}trica, Universidade Federal de Minas Gerais, Av.
%Ant\^{o}nio Carlos 6627, Pampulha, 31270-010, Belo Horizonte, MG,
%BRAZIL. }  % Please supply  
%
%\cortext[a]{Corresponding author}        

% \thanks[footnoteinfo]{ Corresponding author C. A. Maia. Tel. +55-31-3409-3408. Fax. +55-31-3409-4810}         

\begin{abstract}                          % Abstract of not more than 200 words.
Stochastic Event Timing is a fundamental issue in developing both analytic and simulation models for stochastic systems.
Generalized Erlang distributions are quite useful for generating those random events in a quite general way by inserting intermediary states with markovian jumps.  One very important and celebrated generalization of the Erlang distribution  was made by D. R. Cox in the middle 50's. This paper discuss further the Cox generalization and   presents  an even more general topology, capable of representing any practical distribution. As an application, we  revisit  the classical problem of the first  two moments matching, and derive minimal topologies in terms of number of states, then the results are compared with those found in literature. At the end of the paper, we show how the generalized structure can be use for timing  general stochastic discrete-event models for analytic and simulation purposes.
\end{abstract}

                             % keyword list or with the 
                                          % help of the Automatica 
                                          % keyword wizard

\end{frontmatter}

\section{Introduction}   \label{sec:introduction}

 Analytic and computer simulation models  are  well known ways of evaluating performance of  Stochastic Timed Discrete-Event Systems,  as well as Stochastic Hybrid Systems \cite{Cassandras_livro2008}. Exponential distribution is important for timing  events  in such structures. One of remarkable feature of this distribution is the fact that it allows us to construct Markovian models for stochastic process, which can be manipulated analytically with a low computer cost. It is attractive even for Monte Carlo simulation due to its simplicity to be generated from a uniform distribution.

In addition, we can use exponential distribution to represent other more complex distributions by spliting a lifetime of an event into phases, being the jump time between the phases exponentially distributed (markovian jumps). This fantatic fact allow us to convert semi-makovian process into makovian one. A Pioneer in this field was A. K. Erlang with his work in traffic engineering. A rather comprehensive compilation of the family of Erlang distribution can be found in \cite{Bolch2006}.

The applications of the Erlang distributions and their derivations  in  science and engineering is quite vast.  To illustrate this fact, we  list in the following some  results, obviously, without intention to be exaustive: In computer-based medical systems, for modeling deseases  \cite{RebeccaRollins2015}, in realibility engineering, for modeling failure rate \cite{QihongDuan2016}; In electrical systems, for modeling the delay behavior of electromagnetic pulses \cite{Arnaut2016} \cite{QianXu2018} in performance  analysis of antennas;  In communication systems to model personal communication netwoks \cite{FangChlamtac1999} and  to analyse network security \cite{PAgarwal2017}; In Smart Grid Sytems  to construct models for power allocation in Home Area Networks \cite{Li2016}; In Computer systems, for  analysis of computer networks \cite{AnaBusic2010} and software testing \cite{Khomonenko2016}.

Regarding modeling and control  of stochastic process, we can cite: for observation
 distributions \cite{MAdes2000};  in the description of   noisy sampling intervals \cite{BoShen2017}; for  ``markovianization'' of  semi-markov jump linear systems \cite{Jafari2017}  \cite{Li_F_Autom2015};  to model noise in nanosensors  \cite{Soltani2017};  in an application of item processing time for a dynamic pricing problem \cite{ChenL2011}; and in a fault model for  stochastic discrete event system \cite{AmmouraR2017}.
 
In this context, in this paper we are concerned with the fundamental issue of constructing general distributions though markovian jumps. In this sense, we first show that the Cox model (a classical and reputed one that generalize Erlang Distribution) can be made even more general by allowing more flexibility in the topology. We will denote this model as a Generalized Cox Model. With this model,  we  approach the problem of moments matching, which is an important issue in modelling and performance evaluation of stochastic systems \cite{Osogami-2006} \cite{Lagershausen2015} \cite{Brandwajn2019}.   We formulate the problem as a constrained optimization problem and we establish  the least value for the ratio second moment/first moment. Then we prove that we can have even simpler structures to solve the mean and variance matching than the ones presented classically in the literature. Finally, we show how the generalized structure can be used for event timing in stochastic discrete-event system for analytic and computer simulation purposes.  

The sequence of this paper is organized as follow: in Section \ref{section:method}, we present the methodology and the contributions of this paper as well as the comparison with the results found in literature. In Section \ref{section:application} we explain how the generalized structure can be used in the modeling of event generation  for stochastic discrete-event system. The conclusion and perspectives for this work is presented in Section \ref{section:conc}.

\section{Methodology}\label{section:method}
\begin{figure}[h]
\centerline{\includegraphics[width=\columnwidth]{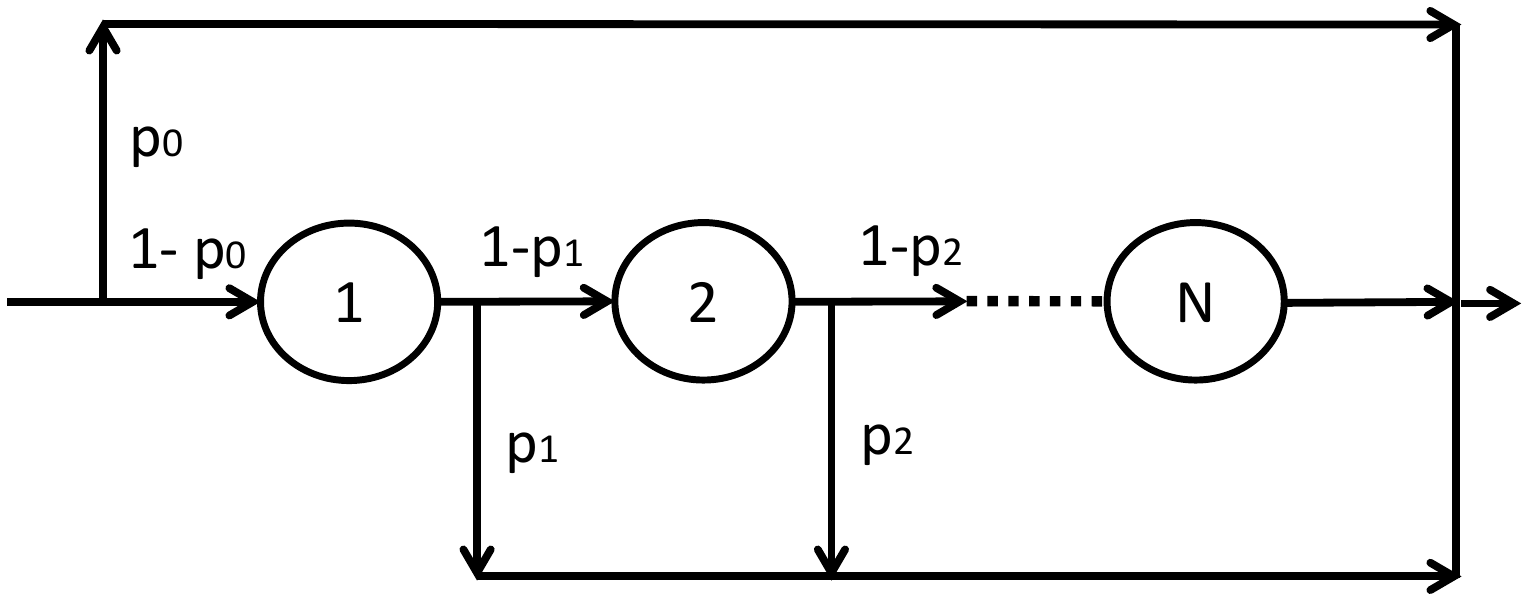}}
\caption{ Cox Distribution with N stages: after staying an exponentially distributed time on a stage $i$ one has a probability  $p_i$ of abandonment and  $1-p_i$ to  go to the next stage.  }
\label{Cox}
\end{figure}

The ideas of  A. K. Erlang was further generalized in order to represent more complex lifetime distribution by  D.R. Cox (Cox \cite{Cox55}). So let us refer to Figure \ref{Cox} to recall how Cox distribution works: initially, a bacteria (or a task or a client in a queue network system)  has a probability $p_0$ of death (or, in practice, negligible service time) and a probability $1-p_0$ of entering the first stage; once in the first stage, it must spend $T_1$ time units exponentially distribute with rate $ \lambda_1$;  after completing this time, it  has a probability $p_1$ of death and a probability $1-p_1 $  of entering into the second stage, and so on. The process repeat till the state $N$ is reached. As a result, the Laplace transform  of the resulting pdf is given by:
\begin{equation}
 f_{c}(s)= q_0+ \sum_{j=1}^{N}\prod_{k=1}^{j} \frac{q_{k}\lambda_{k}}{s+\lambda_{k}},
 \label{fs_Cox}
 \end{equation} 
 \noindent for which $q_{0} =p_{0}$  and  $q_j = p_{j}\prod_{k=0}^{j-1}(1-p_{k}) $ for $j \in \{1, \ldots N \}$,   with a convention that $p_{N}=1$.
 
\begin{figure}[H]
\centerline{\includegraphics[width=\columnwidth]{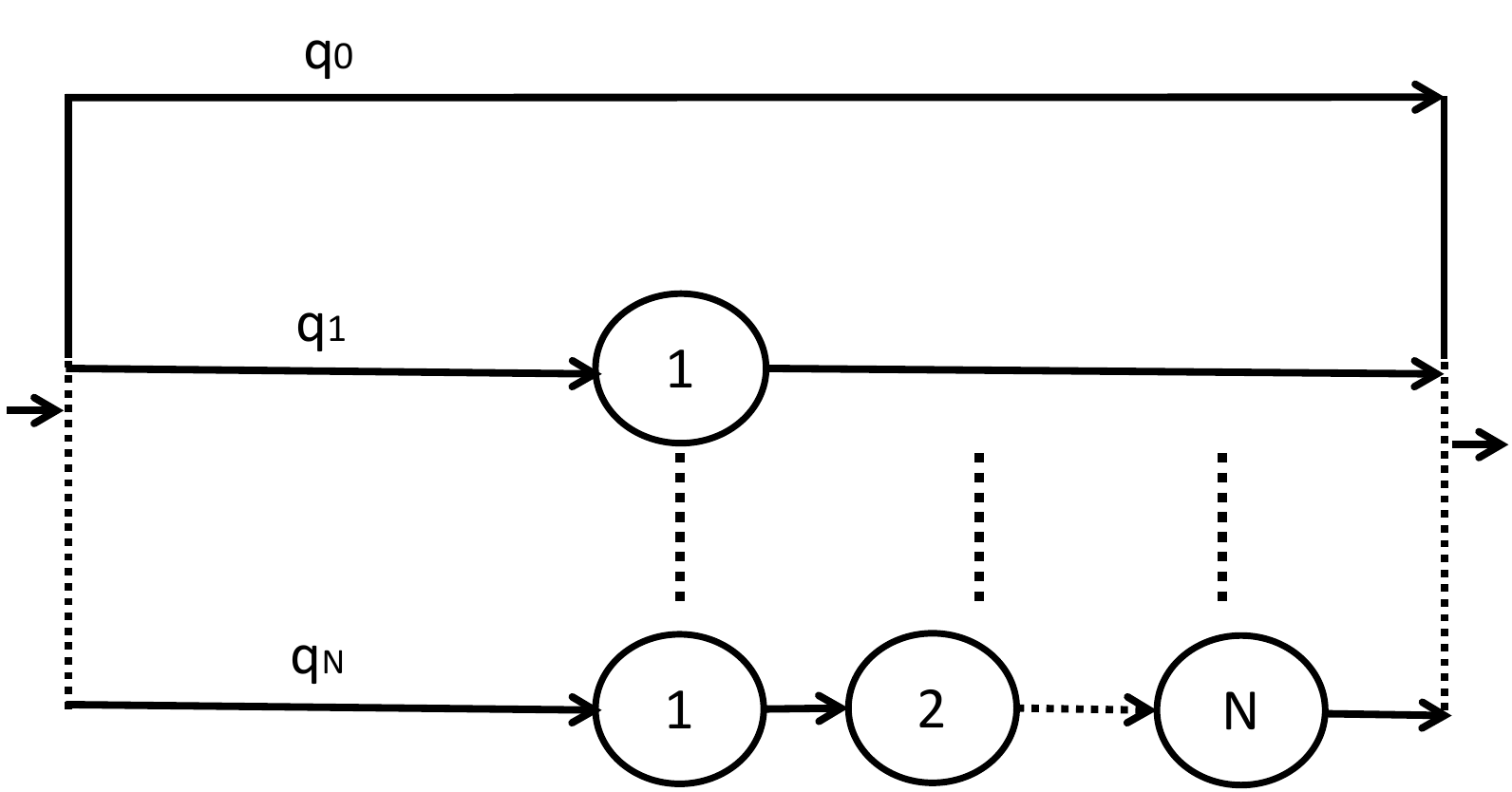}}
\caption{ Equivalent Cox Distribution by reorganizing the layout of the states.  Routing probabilities  are $q_{0} =p_{0}$ and $q_j = p_{j}\prod_{k=0}^{j-1}(1-p_{k}) $ for $j \in \{1, \ldots N \}$,    with a convention that $p_{N}=1$.}
 \label{CoxEquiv}
\end{figure}

\begin{figure}[h] 
\centerline{\includegraphics[width=\columnwidth]{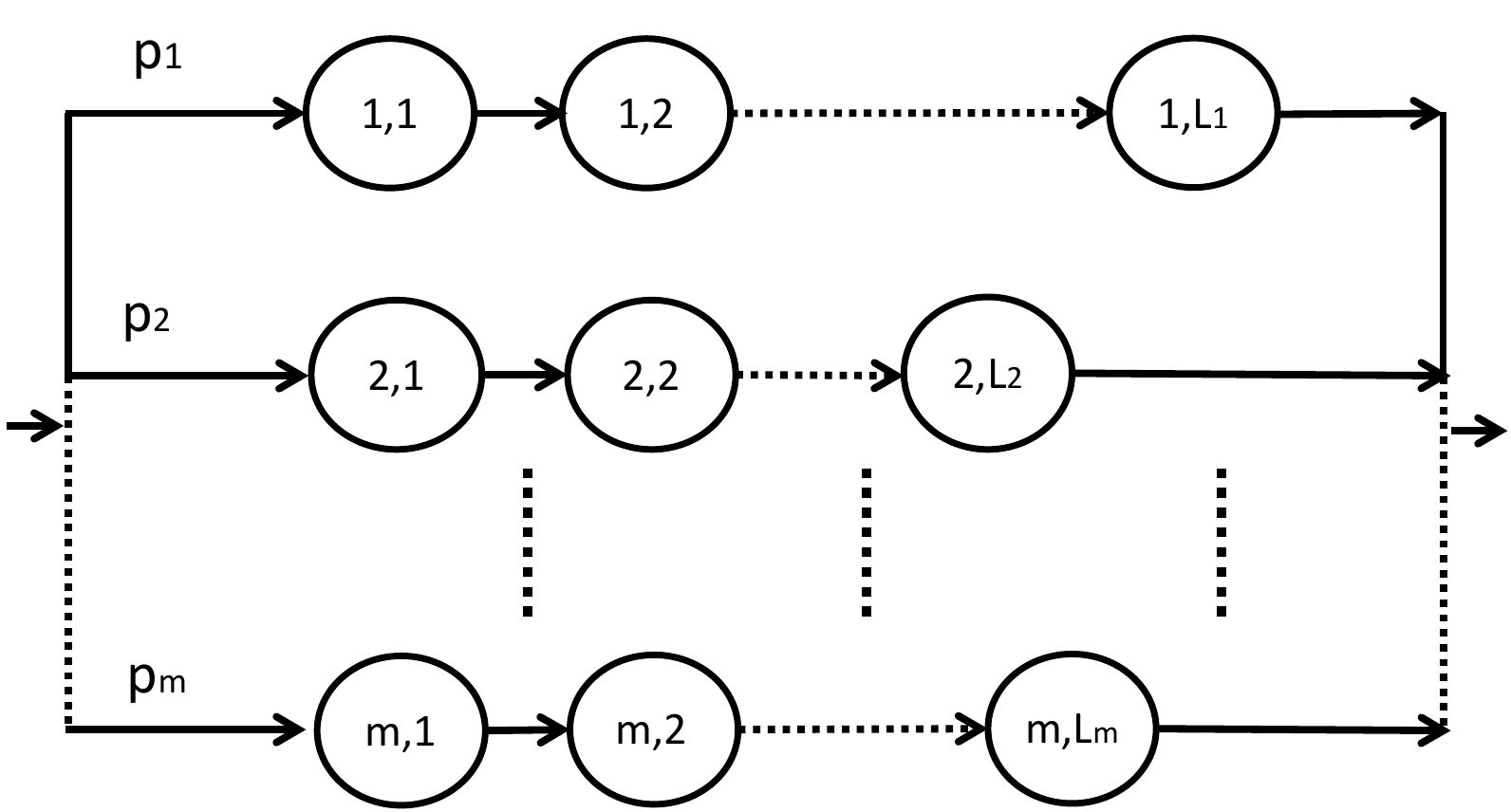}}
\caption{Generalized Cox distribution with $m$ branch routing. Each branch has $L_j$ states for $ j = 1, \ldots m $,   whose  routing probability  is equals to $p_j$.}
\label{CoxGeneralized}
\end{figure}

In the following, we further investigate Cox Distribution Topology. The objectives are two fold: first we present an even more general distribution topology able to generate, for instance, any practical distributions; from this general distribution we derive minimal topologies  enabling us to exactly match the first and second moments (\textit{e.g.} mean and variance). Then we compare the results with those presented in the literature.   
First of all let us recall that  Cox topology, depicted in Figure\ref{Cox}, can be seen as grid composed by branches, which in turn is compose by a sequence of states. In fact, in \cite{Augustin1982}, the authors show that Cox topology can be reorganized equivalently as depicted Figure \ref{CoxEquiv}, being  $q_j$ the routing probabilities, and the time to jump to the following state exponentially distributed.  This equivalent arrangement explicit even more the fact that Cox distribution is obtained by following a sequence of states, that's it, in order to access state $j$, we must mandatory visit all previous states $1,\ldots, j-1$.  
  This very interesting  arrangement  lead us to think  that Cox structure can be  made even more general, by considering independent branches with arbitrary number of states as show in Figure \ref{CoxGeneralized}.
So let us explain this process a little bit more, now in terms of a service time: to generated a random service time that respect  this distribution, which we call hereafter generalized Cox distribution,  we first select randomly a branch $j$, $ j = 1, \ldots m $, with probability  $p_j$. Once a branch is selected,  the process must follow its  assigned sequence of states, considering that the events that cause the transition to the adjacent states  are independent and exponentially generated. For instance, let's say that a branch $j$ is selected with probability  $p_j$, then the process must ``jump''   sequentially from state $1$ till $L_j$, in order complete the service. The resulting distribution  is really quite general.  In fact, let us give an interpretation of this process  in terms of a multi-class client-server system,  for which we can  consider  different service times depending on the type of the client.  In order to do this, we first consider that the jumping rate off state $kj$ denoted as $\lambda_{jk}$.  So  this general distribution can represent situations in which some clients are  served with a negligible service time  with a probability $p_1$ (in this case, $L_1=1$ and $\lambda_{11} \to \infty $, meaning, in practice, that the client pass directly through the server),  other clients are served in a exponential basis with probability $p_2$, others obey a quite more complex way service-time distribution with probability $p_3$, etc.

In the sequel, we show that this generalization of the Cox distribution is really capable of representing more general distribution than the  classical Cox distribution.

\begin{property} \label{proper:GenCox} The pdf of  generalized Cox distribution,  expressed by Laplace transform, given by Equation \ref{fs_genCox}, is more general than the one obtained by the Cox distribution, given by Equation \ref{fs_Cox}.
\begin{equation}
 f_{g}(s)= \sum_{j=1}^{m}\sum_{k=1}^{L_j} \frac{p_{j}\lambda_{jk}}{s+\lambda_{jk}}.
 \label{fs_genCox}
 \end{equation}
\begin{proof} To show that the presented topology , depicted Figure \ref{CoxGeneralized}, is really more general, lets us  give an example of pdf that can be realized by this topology but not with the Cox arrangement.

So we consider $m=2$, $L_1=L_2=1$, and simply denote   $\lambda_{11}= \lambda_1$ and   $\lambda_{22}= \lambda_2$. This consideration lead us to the following Laplace transform:
\begin{equation}
f_{g}(s)= \frac{p_1\lambda_1}{s+ \lambda_1}+ \frac{p_2\lambda_2}{s+ \lambda_2}= \frac{(p_1\lambda_1+p_2\lambda_2)s+ \lambda_2 \lambda_1 }{(s+ \lambda_1)(s+ \lambda_2)},
\end{equation}
\noindent using the fact  that $ p_1 + p_2 =1$. 

Comparing this function with those that we can synthesize by means of a Cox topology, we can see that the only possibility to achieve  the complete matching, \textit{i.e.} $ f_{c}(s)= f_{g}(s)$,  is by  trying to find routing probabilities, let's say  $q_1$ and $q_2$   for a Classical Cox topology in which $N=2$ and $q_0=0$.  For this situation, we have:

\begin{equation}
f_{c}(s)= \frac{q_1\lambda_1}{s+ \lambda_1}+ \frac{q_2\lambda_1\lambda_2}{(s+ \lambda_1)(s+ \lambda_2)}= \frac{q_1\lambda_1s+ \lambda_2 \lambda_1 }{(s+ \lambda_1)(s+ \lambda_2)}.
\end{equation}

As a consequence, the desired equality is only ensured if   $ q_1\lambda_1 = p_1\lambda_1+p_2\lambda_2$, which in turn is only solvable if $ p_1 +  p_2\frac{\lambda_2}{\lambda_1} \leq 1 $. Obviously, this inequality cannot be always true: take for instance $p_1=p_2=0.5$ and $\lambda_2=2\lambda_1$.

\end{proof}

\end{property}

\begin{rema}We have shown so far  that the generalized topology depicted in Figure \ref{CoxGeneralized} is, in fact,  even more general than the classical  one. In fact, we can easily show that this distribution can represent any pdf whose Laplace transform have negative real poles. In this sense, they can be used to approximate any  probability distribution with an arbitrary precision. The main advantage of such approximation is the fact that it allows us to derive analytic Markov models, with can be solved much more faster than Monte Carlo Simulation.
\end{rema}

In the sequence, we  exploit further this this topology by revisiting the classical problem of first two moments matching. We will show some new results concerning minimal achievable moments and minimal topologies.

\subsection{Moment matching problem}

After showing the previous generalization, let us  further investigate the generalized topology in terms of moments matching. For a general pdf expressed as \ref{fs_genCox}, we can show that:

\begin{equation}
f_g(s)= \sum_{k=1}^{\infty}\frac{(-1)^{k} E[T_{g}^{k}]s^k}{k!}.
\end{equation}

So the $k^{th}$ moment of $T_g$ can be computed as:
\begin{equation}\label{Eq_Gen_Mom}
E[T_{g}^{k}]=(-1)^{k}f_{g}^{(k)}(0),
\end{equation}
\noindent in which $f_{g}^{(k)}(0)$ is the $k^{th}$ derivative of $f_{g}(s)$ evaluated at $s=0$.

In this paper,  we are  particularly interested in revisiting the classical problem of first and second method (\textit{e.g.} mean and variance) matching. Formally our problem can be  stated as: \\

\textit{ Find the minimum number of states of topology given in Figure \ref{CoxGeneralized}, as well as their jumping rates,  ensuring that the resulting  distribution has mean $\mu$ and variance $\sigma^2$. } \\

This is a well known problem in literature and we want to shed more light on it, as well as presenting new results. At the end of this paper we will make comparisons with the results found in literature.

Deducing the general expression for the moments using Equation \ref{Eq_Gen_Mom} is a complicated and tedious task for general pdf's. However,  for the first and second moments,  we can exploit properties of our topology in order to simplify the deduction. In fact, the  exponential distribution, with rate $\lambda$, has mean $ 1/\lambda$ and variance $  1/\lambda^2$. So for a given branch $j$ of  the process  depicted in the Figure \ref{CoxGeneralized} we have a sum of independent  random variable exponentially distributed, which we denote by $T_j$,  whose  mean and variance are respectively are given by:

\begin{equation}
 \nonumber E[T_j]=\sum_{k=1}^{L_j} 1/\lambda_{jk}\\
Var[T_j]= \sum_{k=1}^{L_j} 1/\lambda_{jk}^2,
\label{Erlang_series}
\end{equation}
in which $\lambda_{jk}$ is a jump rate off the state $k$ in the branch $j$.

In order to properly write the equations, we denote hereafter $x_{jk}=1/\lambda_{jk}$ and  $T_g$  the random variable that respect generalized Cox distribution, as depicted in Figure \ref{CoxGeneralized}. 

Recalling that  for any random variable, let's say $X$, $E[X^{2}]= E[X]^2 + Var[X] $,  we  deduce that:
\begin{eqnarray}
\label{CoxGen__mean}E[T_g]=\sum_{j=1}^{m}p_j(\sum_{k=1}^{L_j}x_{jk}) \\
\label{CoxGen_2ndMom}E[T_g^2]= \sum_{j=1}^{m}p_j[(\sum_{k=1}^{L_j}x_{jk})^2+(\sum_{i=1}^{L_j}x_{jk}^2)].
\end{eqnarray}
As as result,our problem consists in finding a solution for the following system of equations, while  minimizing the total number of states $(L_1+ \ldots + L_m)$: 
 \begin{eqnarray}
 \nonumber\sum_{j=1}^{m}p_j(\sum_{k=1}^{L_j}x_{jk})= \mu; \\
 \sum_{j=1}^{m}p_j[(\sum_{k=1}^{L_j}x_{jk})^2+(\sum_{i=1}^{L_j}x_{jk}^2)]  = \mu^2+ \sigma^2.
\label{CoxGeneralized_eq}
\end{eqnarray}
 \begin{propo}
 For a given generalized Cox distribution, as depicted in Figure \ref{CoxGeneralized}, the minimum value of  $E[T_g^2]$, in terms of the routing probabilities  and $E[T_g] = \mu$, is given by
 \begin{equation}
  \frac{E[T_g^2]_{mim}}{\mu^2•} = \frac{1}{\sum_{j=1}^{m}\frac{p_jL_j}{1+L_j}}.
\label{fmin_eq1}
\end{equation}

\begin{proof} We look for a minimum value for $E[T_g^2]$, given $E[T_g]= \mu$,   by studying the following  optimization problem: 
\begin{equation}
\begin{aligned}
& \underset{x_{ij}}{\text{Min}}
& &  E[T_g^2] \\
& \text{subject to}
& & \sum_{j=1}^{m}p_j(\sum_{k=1}^{L_j}x_{jk})-\mu =0.
\end{aligned}
\label{Optim_total}
\end{equation}
If an optimum solution exists, the Lagrange multipliers method lead us to the following equations:
\begin{eqnarray}
 \nonumber 2p_j(\sum_{k=1}^{L_j}x_{jk})+ 2p_jx_{jk} - p_j\gamma =0 \\
\nonumber \sum_{j=1}^{m}p_j(\sum_{k=1}^{L_j}x_{jk})= \mu,
\label{Erlang_total}
\end{eqnarray}
\noindent whose $\gamma$ is the Lagrange multiplier. 

Without loss of generality,  suppose that  $p_j \neq 0$, then optimal solution occurs when all $x_{jk}$ are  given by $ \frac{\gamma}{2(1+ L_j)}$. Therefore  we can show that minimum value is such that: 

\begin{eqnarray}
  \nonumber E[T_g^2]_{mim} = (\sum_{j=1}^{m}\frac{p_jL_j}{1+L_j})\frac{\gamma^2}{4} \\
  \mu = (\sum_{j=1}^{m}\frac{p_jL_j}{1+L_j})\frac{\gamma}{2}.
\label{fmin}
\end{eqnarray}

As a consequence we can deduce  that:

\begin{equation}
  \frac{E[T_g^2]_{mim}}{\mu^2•} = \frac{1}{\sum_{j=1}^{m}\frac{p_jL_j}{1+L_j}}
\end{equation}

\end{proof}

\end{propo}

We can obtain an lower bound for this minimum solution by observing the following inequality: 

$$ \sum_{j=1}^{m}\frac{p_jL_j}{1+L_j} \leq \frac{L_{j^{*}}}{1+L_{j^{*}}}$$

in which  $j^{*} = \underset{j}{Arg}~ Max \frac{L_j}{1+L_j}$ . Therefore:

%Proof: Let $j^{*} =  \underset{j}{arg max L_j}$, write $ p_j^{*}= 1- \sum_{ \underset{j \neq j^{*}} {j=1}  }^{m}p_j $. So  
% $ \sum_{j=1}^{m}p_jL_j = L_{j{*}}-  \sum_{ \underset{j \neq j^{*}} {j=1}  }^{m}p_j (L_{j{*}} -L_j)  \leq L_{j{*}} = \underset{j}{Max L_j}. $ 

\begin{equation}
 \frac{E[T_g^2]_{mim}}{\mu^2•}  \geq   \frac {1+L_{j^{*}}}{L_{j^{*}}} = 1+  \frac {1}{L_{j^{*}}}.
\label{ineq_fundam}
\end{equation}

So far we have found a minimum value for the second moment for the Generalized Cox Topology \ref{CoxGeneralized}, given the mean and the routing probabilities. We have  as well  establish an interesting lower bound, based on the number of states, given by Inequality \ref{ineq_fundam}.

\begin{rema}We remark that the lower bound \ref{ineq_fundam} was presented before using a completely different approach in \cite{CommaultPhaseType2003}, as corollary of  a more general result presented before in \cite{Aldos_Shepp1987}. However in those papers, the authors do not  concern  in minimizing second moment, given the mean,  as we did, and do not present a general expression for the minimum value as show in Equation \ref{fmin_eq1}.
\end{rema}

\subsection{Minimal Structures for First and Second  Moment Matching}

\begin{figure}[H]
\centerline{\includegraphics[scale=0.8]{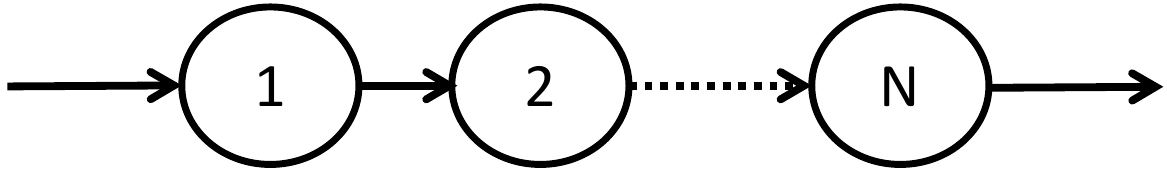}}
\caption{  A particular case: Hypoexponential (or generalized Erlang) distribution }
\label{Hypoexponential}
\end{figure}

The result \ref{ineq_fundam} ensures that the ratio  variance  /mean  of the distribution (we recall that$ Var[T_g^2] = E[T_g^2]  - \mu^2$), can not be smaller than the one obtained for the longest branch. So hereafter we focus on what we can achieve by using just one branch with $N$ states.  The obtained distribution in this case is known as Hypoexponential (or generalized Erlang) distribution, and it is depicted in Figure  \ref{Hypoexponential}. 

In this particular situation, the minimum value for the second moment, is given  by Equation \ref{fmin_eq1}:

\begin{equation}
  \frac{E[T_g^2]_{mim}}{\mu^2•} = \frac{1}{\frac{N}{1+N}}= 1 +\frac{1}{N}.
\label{fmin_Hypo}
\end{equation}

It is well known that this minimum value is achieve for a Erlang Distribuition, with $x_i= \mu/N$ if $\frac{\sigma^2}{\mu^2}= \frac{1}{N}$. So let us investigate more general situations in which this equality is not true.  In this sense, our moment matching problem is simplified as solving the following pair of equation for a minimum $N$:
\begin{eqnarray}
\nonumber \sum_{i=1}^{N} x_i = \mu \\ 
\sum_{i=1}^{N} x_i^2 =  \sigma^2   
\label{Equation_simplif}
\end{eqnarray}
First, we must observe that triangular and internal product inequalities (Cauchy-Schwartz Inequality) lead to:
\begin{equation}
\frac{(\sum_{i=1}^{N} x_i)^2 }{N} \leq \sum_{i=1}^{N} x_i^2 \leq (\sum_{i=1}^{N} x_i)^2 
\label{pob}
\end{equation}

%As a second observation, if a solution exists, then $\sum_{i=1}^{N} x_i^2 \geq \frac{\mu^2}{N}$.  
%In fact, by changing variable as $z_i = x_i - \frac{\mu}{N}$, we have: 
%\begin{equation}
%  x_i = z_i + \frac{\mu}{N}.
%\end{equation}
%
%And, as a consequence, $\sum_{i=1}^{N} z_i = 0$.  Therefore:
%
%\begin{equation}
%\sum_{i=1}^{N} x_i^2 = \sum_{i=1}^{N} (z_i + \frac{\mu}{N})^2 \nonumber 
%\end{equation}
%\begin{equation}
%\sum_{i=1}^{N} x_i^2 = \sum_{i=1}^{N} (z_i^2 + 2 \frac{\mu}{N}z_i + \frac{\mu^2}{N^2}) \nonumber
%\end{equation}
%\begin{equation}
%\sum_{i=1}^{N} x_i^2 = \sum_{i=1}^{N} z_i^2 + \frac{2\mu}{N} \sum_{i=1}^{N} z_i + \sum_{i=1}^{N} \frac{\mu^2}{N^2}
% \end{equation}
%
%Since  $\sum_{i=1}^{N} z_i = 0$ , we ensure that:
%
%\begin{equation}
%\sum_{i=1}^{N} x_i^2 = \sum_{i=1}^{N} z_i^2 + \frac{\mu^2}{N} \geq \frac{\mu^2}{N} 
%\label{sz3}   
%\end{equation}

As a result, provided that a solutions exists, they are all such that:
\begin{equation}
\frac{\mu^2}{N} \leq  \sum_{i=1}^{N} x_i^2 \leq \mu^2
\label{eq1}
\end{equation}
So the number of stages $N$ must satisfies:

\begin{equation}
\frac{\mu^2}{N} \leq \sigma^2 \rightarrow  N\geq \frac{\mu^2}{\sigma^2} .
\label{Nmi}
\end{equation}

Since $N$ is an integer, its smallest possible value is \footnote{$ \lceil x \rceil$ stands for the ceil of $x$, \textit{i.e.} smallest integer greater than or equal to $x$.} $N= \lceil \frac{\mu^2}{\sigma^2} \rceil $. \\

Keeping in mind those observations, we  present a solution for the system of equations \ref{Equation_simplif}. First we observe that if $ \frac{\mu^2}{\sigma^2} =1 $, the solution is trivial, with only one state, that is $N=1$. So  let us concentrate our attention in the situations for which  $ \lceil \frac{\mu^2}{\sigma^2} \rceil \geq 2$.

\begin{propo}[Almost Erlang Solution] \label{propoErlangGen} If $ \lceil \frac{\mu^2}{\sigma^2} \rceil \geq 2$, a solution for the system of equations \ref{Equation_simplif} is given by:
\begin{equation}
x_j =  \frac{\mu}{N} - \frac{\alpha_N}{\sqrt{(N-1)}} ~~( 1 \leq j \leq N-1), 
\label{solut_til_n-2}
\end{equation}
 and

\begin{equation}
 x_N = \frac{\mu}{N} +\sqrt{(N-1)} \alpha_N.
\label{solut_n}
\end{equation}
being $N= \lceil \frac{\mu^2}{\sigma^2} \rceil $,  $\alpha_N = \frac{\sqrt{N\sigma^2 -\mu^2}}{N}$.\\

\begin{proof}
By denoting $x_j=y$ $( 1 \leq j \leq N-1)$   and  $x_N = z$, it is straightforward   to check  that :

%\begin{equation}
%\left\{\begin{matrix}
%(N-1)y + z= \mu,
%\label{EqXnXn1} & \\ 
%(N-1)y^2 +z^2 = \sigma^2 & 
%\end{matrix}\right,
%\end{equation}

\begin{equation}
\left\{\begin{matrix}
(N-1)y + z= \mu,
\label{EqXnXn1} & \\ 
(N-1)y^2 +z^2 = \sigma^2, & 
\end{matrix}\right.
\end{equation}

which obviously ensure \ref{Equation_simplif}.

\end{proof}
 
\end{propo}

\begin{corollary}[Erlang Solution] In particular, If  $  \frac{\mu^2}{\sigma^2} =N $, a solution is given by $x_i=\frac{\mu}{N}$.
\begin{proof} In this situation, we can verify that $\alpha_N =0$, ensuring the claimed result.

\end{proof}
\end{corollary}

\begin{rema}We remark that the result given by  Proposition \ref{propoErlangGen} is the same as the one obtained by the Erlang Distribution  if  $  \frac{\mu^2}{\sigma^2} =N$.  If it is not the case,  it is important to recall that Erlang distribution does not ensure the first two moments matching. However we prove that is still possible  to  achieve the matching by appropriately changing  the rates of stages. 
\end{rema}

So far, we have established that  if $\sigma^2 \leq \mu^2$, the minimum number of states is given by $ N= \lceil \frac{\mu^2}{\sigma^2} \rceil$, being the means between states provided by Proposition \ref{propoErlangGen}.

\begin{figure}[H]
\centerline{\includegraphics[scale=0.8]{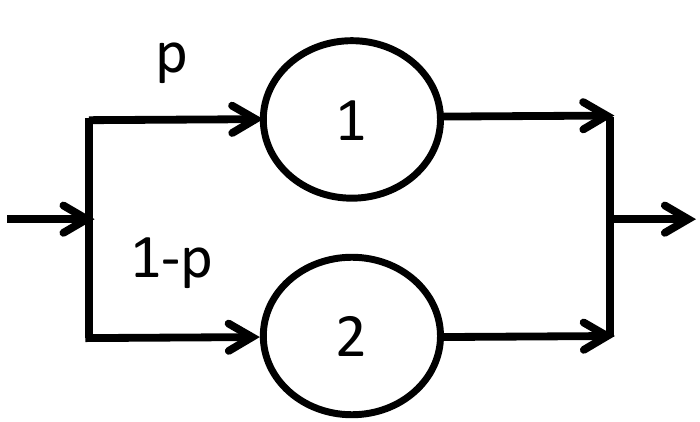}}
\caption{ A particular case: Hyper-exponential  Distribution}
\label{Hyperexponential}
\end{figure}

It remains to solve the cases for which if $\sigma^2$ is strictly greater than  $ \mu^2$. We  address this problem by considering two states and two routing probability, as depicted in Figure \ref{Hyperexponential}.  This configuration, lead us to an Hyperexponential distribution.  In fact, Equation \ref{fmin_eq1} ensures that:
 
 \begin{equation}
 \frac{E[T_g^2]_{mim}}{\mu^2•} = \frac{1}{\frac{p}{1+1}+ \frac{1-p}{1+1}} = 2,
 \end{equation}
 
 which is compatible with our aim to have $ E[T_g^2] = \sigma^2 +\mu^2 \geq  2\mu^2.$

\begin{figure}[H]
\centerline{\includegraphics[scale=0.8]{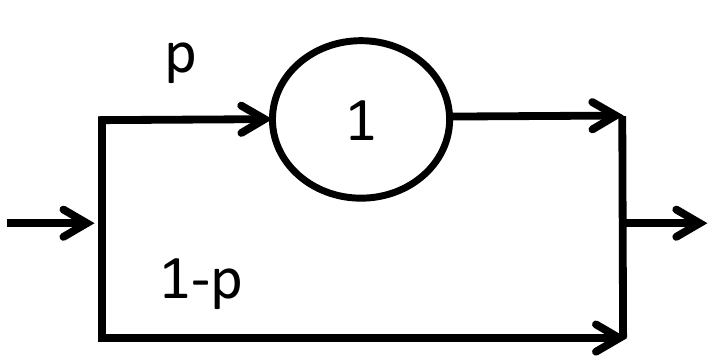}}
\caption{ The simplest form: Hyper-exponential  Distribution}
\label{Hyperexponential-Simplest}
\end{figure}

For this particular structure, the  general system of equation \ref{CoxGeneralized_eq} is reduced to:
\begin{eqnarray}
\label{CoxGen_Cv_geq1_media} E[T_g]=  px_1 +(1-p)x_2 =\mu,\\
\label{CoxGen_Cv_geq1_var} E[T_g^2]=  2px_{1}^2+2(1-p)x_{2}^2=\sigma^2 + \mu^2.
\end{eqnarray} 
Considering, without loss of generality that $x_1 \geq x_2$, we solve the system of equation. The obtained solutions are given by:
\begin{eqnarray}
x_1= \mu (1+ \frac{\alpha}{p}),\\
x_2= \mu (1-\frac{\alpha}{1-p}).
\end{eqnarray}
being $ 0 < p \leq \frac{2}{1 +C_{v}^2}$, $\alpha = \sqrt{\frac{p(1-p)(C_{v}^2-1)}{2}}$ and $C_{v}$ the coefficient of variation, \textit{i.e} $C_{v}= \frac{\sigma}{\mu}$.
So the simplest topology to ensure the desired results is given by choosing $p = p_{max}= \frac{2}{1 +C_{v}^2}$, leading to $x_1= \mu\frac{C_{v}^2 +1}{2}$ and $x_2=0$. This topology, which has only one state, is shown in Figure \ref{Hyperexponential-Simplest}.
\begin{rema}An interpretation of the resulting distribution for this topology, in terms of service time, is the following:
with probability $p$, some clients are served with exponentially distributed service time with mean $x_1$,  while others are served with  negligible service mean with probability $(1-p)$. 
\end{rema}

\textbf{Complete Method Summary \\}

\begin{itemize}
\item  if $\sigma^2 < \mu^2$:  $N= \lceil \frac{\mu^2}{\sigma^2} \rceil $ \\
\begin{figure}[H]
\centerline{\includegraphics[scale=0.8]{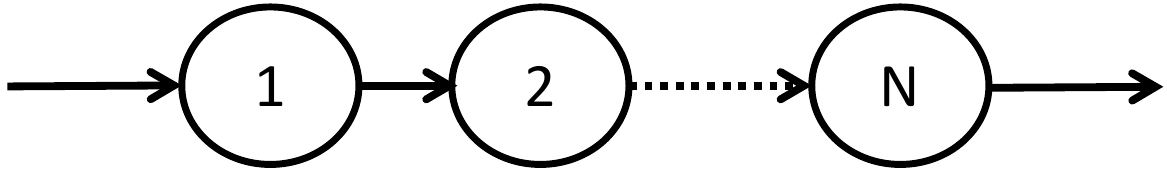}}
\end{figure}
\begin{eqnarray}
 \nonumber  \frac{1}{\lambda_j} & = & x_j =  \frac{\mu}{N} - \frac{\alpha_N}{\sqrt{(N-1)}} ~~~ (1 \leq j \leq N-1), \\
 \nonumber  \frac{1}{ \lambda_{N}} & = & \frac{\mu}{N} +\sqrt{(N-1)} \alpha_N. \\
 \end{eqnarray}
being $\alpha_N = \frac{\sqrt{N\sigma^2 -\mu^2}}{N}$ . \\
\item  if $\sigma^2 \geq \mu^2$:
\begin{figure}[H]
\centerline{\includegraphics[scale=0.8]{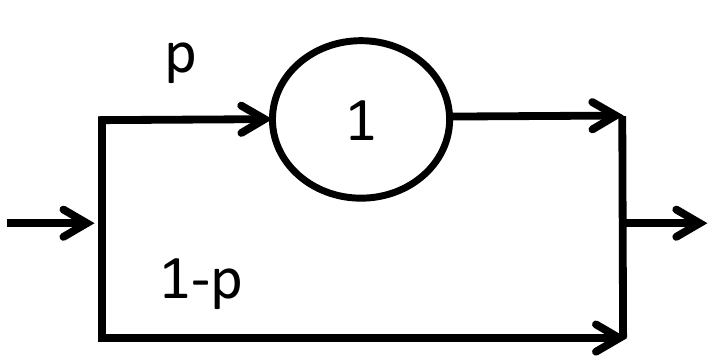}}
\end{figure} 
\begin{eqnarray}
 \nonumber   p= \frac{2}{1 + (\frac{\sigma}{\mu})^2}, ~~~~~~ x_1= \mu\frac{(\frac{\sigma}{\mu})^2 +1}{2}. 
\end{eqnarray}
\end{itemize}
\subsection{Comparison with other results}

The so classical approaches to solve the  model matching problem are presented in \cite{SauerChandy1975} and \cite{Marie1980}. In \cite{SauerChandy1975} the authors presents two topologies which are particular cases of the general Cox distribution presented in Figure \ref{CoxGeneralized}: for  $\sigma^2 \leq \mu^2$, the resulting structure is not minimal since it uses the same number of states, but with the need of a routing probability after the first state;  on the other hand if $\sigma^2 > \mu^2$,  the topology is not minimal as well since it uses two states with means $x_1=\frac{\mu}{2p}$ and $x_2=\frac{\mu}{2(1-p)}$, leading to a particular routing probability $p =\frac{C_{v}^2+1 -\sqrt{ C_{v}^4-1}}{2(C_{v}^2+1)}$. In \cite{Marie1980}, a particular Cox topology is presented for solving the problem for any coefficient of variation. For the cases in which $\sigma^2 < \mu^2$, the resulting structure is the same as proposed by \cite{SauerChandy1975}; if $\sigma^2 > \mu^2$, the structure presents two states with a routing probability after the first state, which is not minimal as well.

\section{Application: Timing Sthocastic Discrete-Event Models}\label{section:application}
\begin{figure}[h] 
\centerline{\includegraphics[width=\columnwidth]{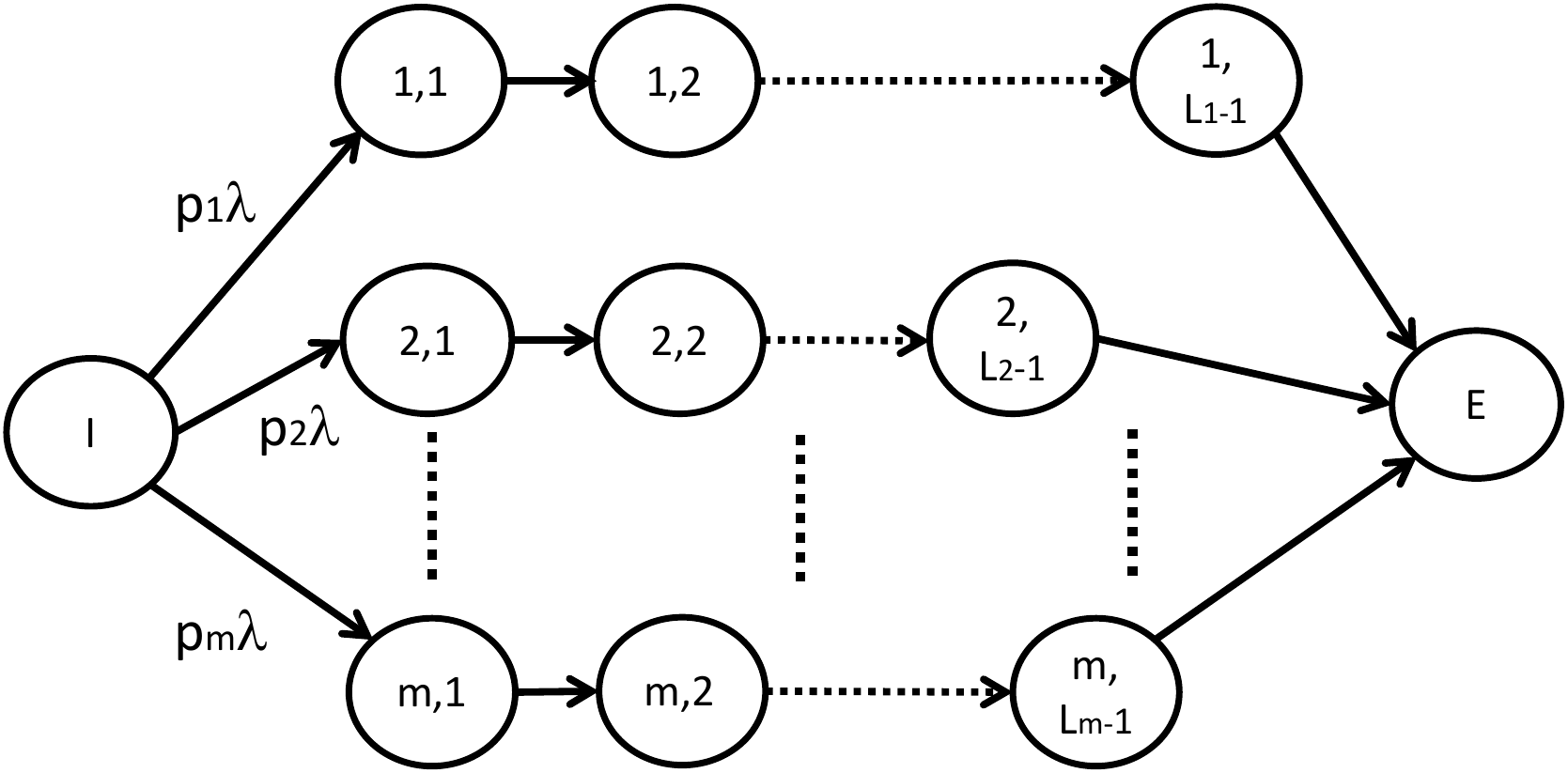}}
\caption{Generalized Cox distribution as an absorbing continuous Markov Chain, in which states I and E represent respectively  the activation and the occurrence of an event.}
\label{CGenMarkov}
\end{figure}

The generalized Cox distribution can be used for timing events in discrete-event system in a quite general way. If we are interested in developing analytic model,  we can observe that
the topology depicted in Figure \ref{CoxGeneralized} can be rewritten as continuous-time Markov chain as show in Figure \ref{CGenMarkov}, whose states  I and E represent respectively and event activation and E its effectively  occurrence, being  value $\lambda$ an infinity (huge)  rate introduced  in order to simulate the routing mechanism. In practice, in order to have a good approximation this rate must be chosen much lager than the others present in the chain.  

So between states ``S'' and ``E'' we have inserted several intermediary states whose transitions operates with markovian jumps in such a way that the `` big  jump''   between ``S'' and ``E''  respects the desired distribution.  In practical situations, minimal topology can be derived by matching the mean and variance of the collected data as show in previous section.

On the other hand, if we are interested computer-simulation models, the generalized Cox distribution samples can be generated from uniform a distribution using the inverse transform method. Suppose that we want to generate a random event-time (for stochastic automata) or a random delay (for stochastic Petri nets), let's say $T_g$. If this random variable is purely exponential with rate $\lambda$, it's well known that it can be generated from an uniform random variable $U \in [ 0~1] $ as:
\begin{equation}\label{Exp_Generation}
T_g= -\dfrac{1}{\lambda}\ln U,
\end{equation}
\noindent ensuring, obviously, that $U \neq 0$.

If $T_g$  respects the  generalized Cox-distribution, as depicted in Figure \ref{CoxGeneralized},   it can be generated by following simple steps: \\

\begin{itemize}
\item Sample  ~$M \in \{1, \ldots m \}$ ~ s.t.~ $P[M=j]=p_j$; \\
\item Sample   $U_k \in [ 0 ~~1],  ~ \forall ~ k \in \{1, \ldots L_M \}$;  \\
\item  Compute $T_g = -\sum_{k=1}^{L_M} \dfrac{1}{\lambda_{Mk}}\ln U_k $,
\end{itemize}

\noindent whose $U_k$ are uniformly distributed random variables in $[ 0 ~1]$ such that $U_k \neq 0$.

In the same way as for analytic models, minimal topology can be derived by matching the mean and variance of the collected data as show in previous section.

\section{Conclusion}\label{section:conc}

With this paper we have expected to contribute to the fundamental problem of  stochastic event timimg by means of markovian jumps, which has a direct impact on the developing analytic and simulation models for Stochastic Systems. To this end we have revisited the classical Cox topology and its equivalent representation, then we derive  an even more general topology, capable of representing quite complex distributions. From this general distribution, we revisited   and reformulated the problem of moments matching in quest of minimal topologies. First we have established an expression for the minimum value of the second moment, given the first one, as well as a lower bound for it. In the sequence, we have presented minimal structures for the first two moments matching problem, which are  simpler  than the ones found in literature. As quite direct application of the results, we show how to generate random events for stochastic discrete-event for analytic and simulation purposes.
For future works, we suggest further investigation concerning higher order moments matching or other metrics concerning transfer function matching.

\bibliographystyle{plain}        % Include this if you use bibtex 

\end{document}